\documentclass{aa}  
\usepackage{graphicx}
\usepackage{txfonts}
\usepackage{natbib}
\usepackage{color}
\usepackage{hyperref}
\hypersetup{colorlinks=True,citecolor=blue,allcolors=blue}
\usepackage{tabularx}
\usepackage{tabu}
\usepackage{float}
\usepackage{xcolor}
\pdfoutput=1 

\begin{document}
   \title{Revealing the inner workings of the lensed quasar \mbox{SDSS J1339+1310}: Insights from microlensing analysis}
\author{C. Fian\inst{1,2}, J. A. Mu\~noz\inst{1,2}, J. Jim\'enez-Vicente\inst{3,4}, E. Mediavilla\inst{5,6}, D. Chelouche\inst{7,8}, S. Kaspi\inst{9}, R. Forés-Toribio\inst{1,2}}
\institute{Departamento de Astronom\'{i}a y Astrof\'{i}sica, Universidad de Valencia, E-46100 Burjassot, Valencia, Spain; \email{carina.fian@uv.es} \and Observatorio Astron\'{o}mico, Universidad de Valencia, E-46980 Paterna, Valencia, Spain \and Departamento de F\'{\i}sica Te\'orica y del Cosmos, Universidad de Granada, Campus de Fuentenueva, 18071 Granada, Spain \and Instituto Carlos I de F\'{\i}sica Te\'orica y Computacional, Universidad de Granada, 18071 Granada, Spain \and Instituto de Astrof\'{\i}sica de Canarias, V\'{\i}a L\'actea S/N, La Laguna 38200, Tenerife, Spain \and Departamento de Astrof\'{\i}sica, Universidad de la Laguna, La Laguna 38200, Tenerife, Spain \and Department of Physics, Faculty of Natural Sciences, University of Haifa,
Haifa 3498838, Israel \and Haifa Research Center for Theoretical Physics and Astrophysics, University of Haifa,
Haifa 3498838, Israel \and School of Physics and Astronomy and Wise Observatory, Raymond and Beverly Sackler Faculty of Exact Sciences, Tel-Aviv University, Tel-Aviv, Israel}
 

  \abstract
  {}
   {We aim to unveil the structure of the continuum and broad-emission line (BEL) emitting regions in the gravitationally lensed quasar SDSS J1339+1310 by examining the distinct signatures of microlensing present in this system. Our study involves a comprehensive analysis of ten years (2009--2019) of photometric monitoring data and seven spectroscopic observations acquired between 2007 and 2017.}
   {This work focuses on the pronounced deformations in the BEL profiles between images A and B, alongside the chromatic changes in their adjacent continua and the striking microlensing variability observed in the $r$-band light curves. We employed a statistical model to quantify the distribution and impact of microlensing magnifications and utilized a Bayesian approach to estimate the dimensions of various emission regions within the quasar. To establish a baseline relatively free of microlensing effects, we used the cores of the emission lines as a reference.}
   {The analysis of the $r$-band light curves reveals substantial microlensing variability in the rest-frame UV continuum, suggesting that image B is amplified relative to image A by a factor of up to six. This finding is corroborated by pronounced microlensing-induced distortions in all studied BEL profiles (Ly$\alpha$, Si IV, C IV, C III], and Mg II), especially a prominent magnification of image B's red wing. These microlensing signals surpass those typically observed in lensed quasars, and the asymmetric line profile deformations imply an anisotropic broad-line region (BLR). We estimated the average dimensions of the BLR to be notably smaller than usual: the region emitting the blue wings measures $R_{1/2} = 11.5 \pm 1.7$ light-days, while the red wings originate from a more compact area of $R_{1/2} = 2.9\pm0.6$ light-days. From the photometric monitoring data, we inferred that the region emitting the $r$-band is $R_{1/2} = 2.2\pm0.3$ light-days across. Furthermore, by assessing the gravitational redshift of the UV Fe III blend and combining it with the blend's microlensing-based size estimate, we calculated the central SMBH's mass to be $M_{BH} \sim2 \times 10^8 M_\odot$.}
   {}


\titlerunning{Structure of the accretion disk and BLR in SDSS J1339+1310}
\authorrunning{Fian et al.} 
\maketitle
\section{Introduction} 
Emission regions in distant active galactic nuclei (AGN) remain inaccessible with the current capabilities of existing telescopes owing to their small dimensions and the vast distances involved. Microlensing variability induced by stellar mass objects in the lens galaxy allows us to infer the sizes of these regions indirectly, thereby overcoming the limitations of direct observations. Microlensing particularly affects compact sources such as the X-ray emitting regions, the accretion disk, and the innermost broad-line emitting clouds. This phenomenon has emerged as an exceptionally powerful tool for probing the structure of lensed quasars (\citealt{Jimenez2012,Jimenez2014,Guerras2013,Guerras2013iron,Motta2012,Motta2017,Munoz2016,Rojas2020}). Specifically, the optical continuum light curves of lensed quasars at redshifts of $z\sim1-2$ can yield measurements of the radii for ultraviolet (UV) continuum-emitting sources surrounding the central supermassive black holes (SMBH; see \citealt{Cornachione2020,Fian2016,Fian2018AD,Fian2021AD,Shalyapin2021,Rivera2023}). Additionally, the wings of broad-emission lines (BELs) observed in quasar rest-frame UV spectra offer valuable insights into the structure and kinematics of the broad-line regions (BLR), as detailed in \citet{Hutsemekers2021}, \citet{Hutsemekers2023}, \citet{Fian2018blr,Fian2021BLR,Fian2023_0957,Fian2023_1004}, and \citet{Popovic2020}. Unfortunately, many lensed quasars either lack sufficient high-quality data or only exhibit weak microlensing effects in their spectra and light curves. The doubly-lensed quasar \mbox{SDSS J1339+1310}, discovered by the Sloan Digital Sky Survey (\citealt{Inada2009}), is a notable exception. It is often referred to as a `microlensing factory' (\citealt{Shalyapin2014,Goicoechea2016}) owing to its pronounced microlensing variations on short timescales in the optical light curves and substantial spectral distortions in several emission lines. The two optically bright images A and B of this quasar, with magnitudes of approximately $r\sim18-19$, are separated by 1\farcs7 and are located at a redshift of $z_s = 2.231$ (\citealt{Shalyapin2021}). An early-type galaxy G at redshift $z_l = 0.607$ (\citealt{Shalyapin2014,Goicoechea2016}) is located $0.\arcsec63$ from image B (\citealt{Inada2009}) and serves as gravitational lens. It is believed that stellar mass objects within galaxy G act as microlenses and strongly affect image B (\citealt{Shalyapin2014}). Systems that exhibit highly variable microlensing events, like this one, are excellent tools for in-depth studies of the central engines of AGN and represent the best targets for subsequent follow-up investigations using suitable facilities.\\
\begin{table*}[h]
\tabcolsep=0.45cm
\renewcommand{\arraystretch}{1}
\caption{Spectroscopic data.}
\begin{tabular}{cccccc} \hline \hline \vspace*{-3mm}\\
Epoch & Date & Image & BEL & Facility & Reference\\ \hline \vspace*{-3mm} \\
1 & 13-05-2007 & A, B & C IV & UH88 & \citealt{Inada2009}\\ 
2a & 16-03-2012 & A & Ly$\alpha$, Si IV, C IV, C III], Mg II & SDSS & \citealt{Dawson2013} \\
2b & 29-03-2012 & B & Ly$\alpha$, Si IV, C IV, C III] & SDSS & \citealt{Dawson2013} \\
3 & 13-04-2013 & A, B & C IV, C III], Mg II & GTC & \citealt{Shalyapin2014} \\
4 & 27-03-2014 & A, B & C IV, C III], Mg II & GTC & \citealt{Goicoechea2016} \\ 
5 & 20-05-2014 & A, B & Ly$\alpha$, Si IV, C III], C IV & GTC & \citealt{Goicoechea2016} \\ 
6 & 18-02-2016 & A, B & C IV & HST & \citealt{Lusso2018} \\ 
7 & 06-04-2017 & A, B & Ly$\alpha$, Si IV, C IV, C III], Mg II & VLT & \citealt{Shalyapin2021}\\ \hline \vspace*{-2.5mm}
\end{tabular}

\small \textbf{Notes.} UH88: University of Hawaii 88\arcsec\ Telescope; 
GTC: Gran Telescopio CANARIAS;
HST: Hubble Space Telescope; 
SDSS: Sloan Digital Sky Survey; 
VLT: Very Large Telescope.
\normalsize
\label{data}    
\end{table*}

In this paper, we conduct a thorough analysis of microlensing signals in both the wavelength and time domains for \mbox{SDSS J1339+1310} to gain deeper insights into the quasar's inner workings. In Section \ref{2}, we show the photometric and spectroscopic data compiled from various literature sources. \mbox{Section \ref{3}} is dedicated to the analysis and interpretation of the data. Section \ref{4} describes our microlensing simulations and explains the Bayesian methods we use to infer the sizes of emitting regions. In Section \ref{5}, we present and discuss our results, encompassing the structure of the continuum- and broad-line-emitting regions, as well as the mass estimation of the central SMBH. Our conclusions are summarized in Section \ref{6}. Throughout this work we assume the cosmological parameters $\Omega_m = 0.3$, $\Omega_\Lambda = 0.7$, and $H_0 = 72$ km s$^{-1}$ Mpc$^{-1}$.

\section{Data and observations}\label{2}
\begin{figure*}
\centering
\includegraphics[width=0.99\textwidth]{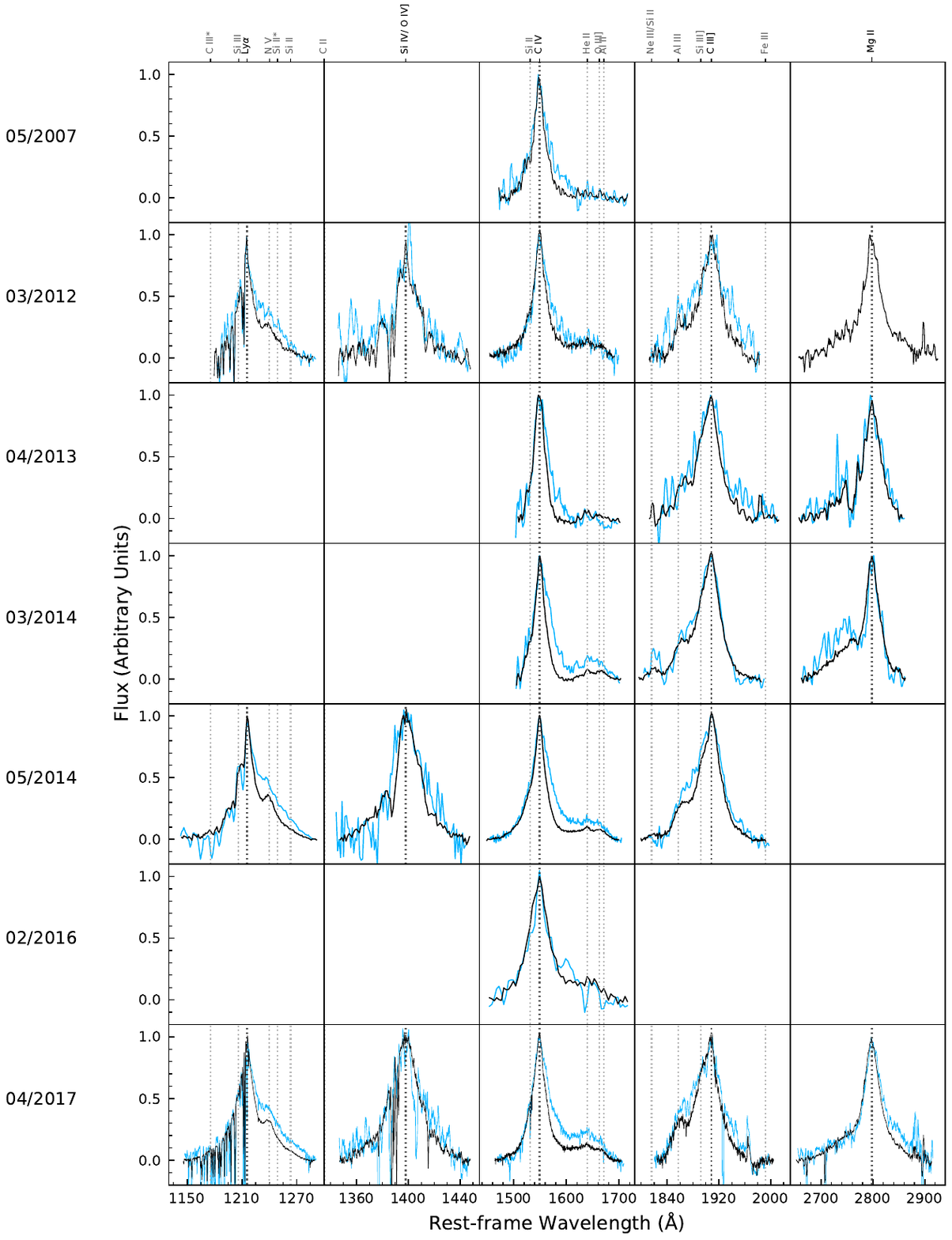}
\caption{Emission line profiles of images A (black) and B (blue) at seven different epochs in the rest-frame after the continuum has been subtracted and the line cores have been normalized. The observations reveal significant deformations in the red wings of image B for all emission lines.} 
\label{emissionlines}
\end{figure*}

\begin{figure*}
\centering
\includegraphics[width=0.99\textwidth]{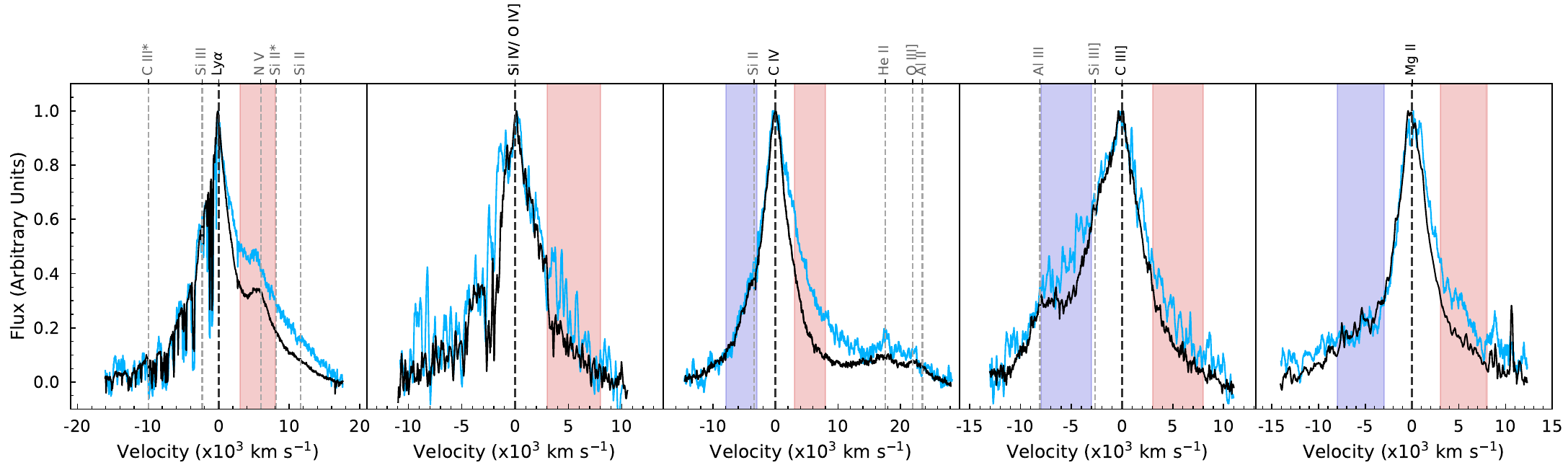}
\caption{
Average Ly$\alpha$, Si IV, C IV, C III], and Mg II emission line profiles (from left to right) of the images A (black) and B (blue). Blue and red shaded regions indicate the selected windows for magnitude difference calculations, spanning $3000-8000$ km s$^{-1}$. We note that the blue wings of Ly$\alpha$ and Si IV are affected by absorption features and have been excluded from our analysis. The observations reveal significant microlensing-induced differences between the images in the red wings of all emission lines. The x-axis is represented on a velocity scale.\\ 
}
\label{mean}
\centering
\vspace*{5mm}\includegraphics[width=0.99\textwidth]{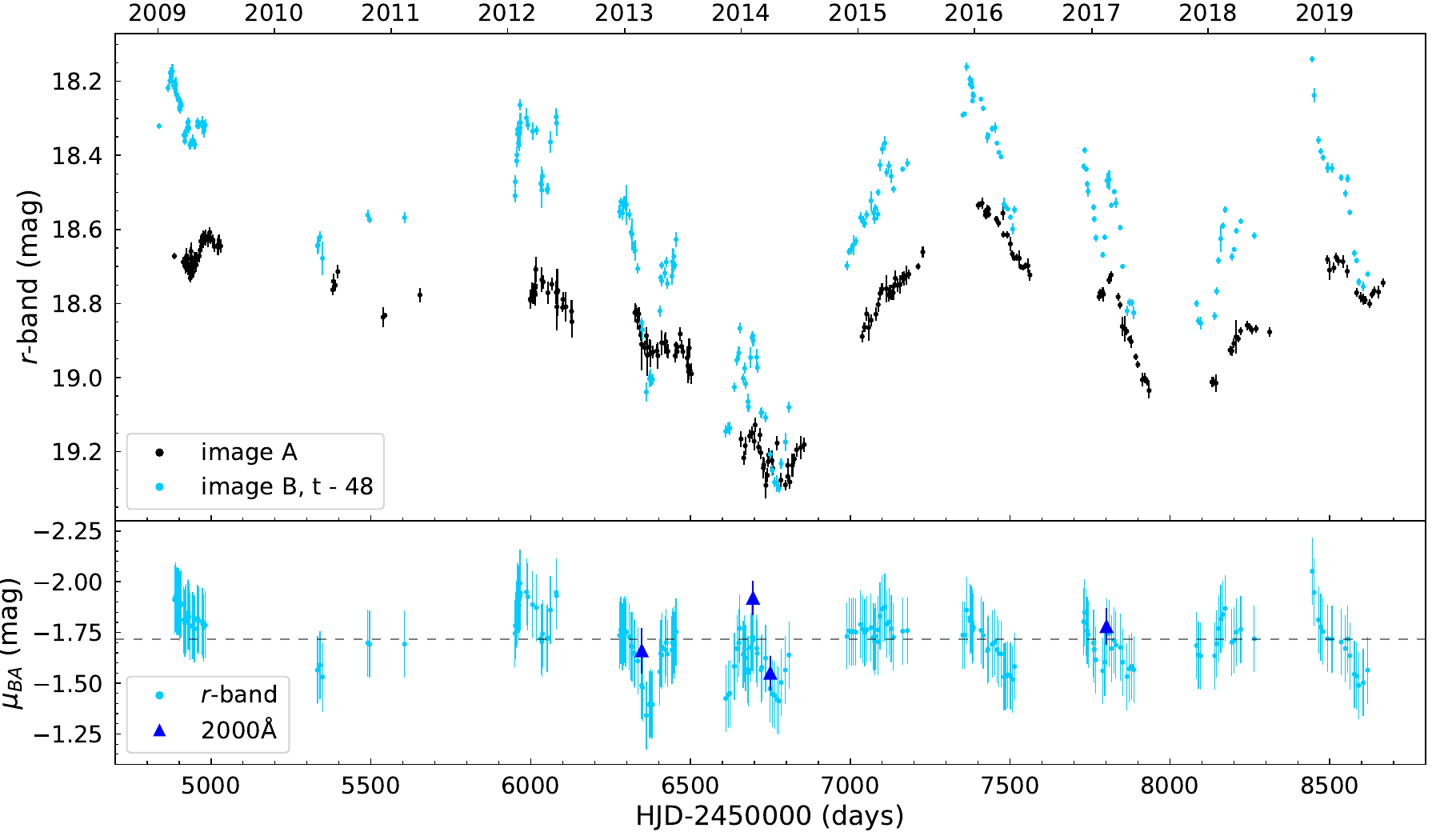}
\caption{Light curves and differential microlensing variability of the quasar SDSS J1339+1310. \textit{Upper panel:} Light curves of the lensed images A (black) and B (light blue), as obtained by the GLENDAMA project (refer to \citealt{Gil-Merino2018}). The light curve of image B is shifted by its time delay ($-48$ days) relative to image A.
\textit{Lower panel:} Differential microlensing variability in light curve B, compared against a polynomial fit to light curve A. The differential microlensing variability has been corrected for the magnitude difference between the images, using the Mg II line core flux ratio as a baseline for no microlensing. The differential microlensing measurements in the continuum at rest-frame 2000\AA, chosen for its close wavelength proximity to the $r$-band at 1930\AA, are depicted as blue triangles for comparison. We note that one of the five available spectroscopic epochs has been discarded due to excessive noise in the data. 
The horizontal axes display Julian (bottom) and Gregorian (top) dates.}
\label{lightcurves}
\end{figure*}

From the literature, we have compiled a set of seven rest-frame UV spectra for images A and B of the gravitationally lensed quasar SDSS J1339+1310, covering a decade from May 2007 to April 2017. These spectra encompass a range of high- and low-ionization lines typical in quasars, such as Ly$\alpha$ $\lambda$1216, \mbox{Si IV $\lambda$1397}, C IV $\lambda$1549, C III] $\lambda$1909, and Mg II $\lambda$2798. The data, which have been fully reduced, are detailed in \mbox{Table \ref{data}}, along with information about the observations and corresponding references. In Figure \ref{emissionlines}, we display superpositions of the emission line profiles for Ly$\alpha$, Si IV, C IV, C III], and \mbox{Mg II} corresponding to images A and B across various epochs. 
In \mbox{Figure \ref{mean}}, we present the average profiles of these BELs for each lensed image. Individual epochs exhibiting noisy data within the emission line range were omitted before calculating the average spectra. We want to emphasize that in Figures \ref{emissionlines} and \ref{mean}, the emission line profile differences are highlighted by subtracting a linear fit of the continuum and normalizing the spectra to the peak of each respective emission line.\\

A visual inspection of the data reveals significant deformations in the emission line wings when comparing images A and B. While the blue wings of the high-ionization lines coincide well with each other, a pronounced enhancement is evident in the red wing of B compared to A across all emission lines. These observed anomalies are likely attributable to microlensing and indicate that the broad-line emission region has an anisotropic structure, as supported by findings in previous studies (e.g., \citealt{Schneider1990, Abajas2002, Lewis2004,Sluse2012,Fian2018blr,Fian2021BLR,Fian2023_1004}). We also observe time-varying distortions in the shelf-like feature located blueward of He II (around $\sim \lambda$1610), manifesting with varying intensities across different epochs. The origin of this feature remains uncertain; potential explanations include an extreme C IV red wing or \mbox{He II} blue wing, or possibly an unidentified species as discussed in \citet{Fine2010} and references therein. The blue wings of the low-ionization lines \mbox{C III]} and Mg II are contaminated by other emission features -- Al III and Si III] in the case of C III], and the `small blue bump' formed by the blending of various iron lines in the case of Mg II -- both exhibiting indications of microlensing.\\ 

Following the discovery of SDSS J1339+1310 (\citealt{Inada2009}), the system was also monitored in the $r$-band within the framework of the Gravitational LENses and DArk MAtter (GLENDAMA) project\footnote{https://grupos.unican.es/glendama/} (\citealt{Gil-Merino2018}). The photometric monitoring was conducted with the 2.0m Liverpool Telescope from 2009 to 2019, comprising 241 observing epochs. The light curves, as shown in the upper panel of Figure \ref{lightcurves}, are publicly available on the GLENDAMA website. For a comprehensive analysis of the data, we refer the reader to the detailed study conducted by \citet{Shalyapin2021}. 

Lensed quasar images exhibit a time delay in their arrival as they traverse different paths through the universe. \citet{Shalyapin2021} determined the time delay between the images to be $\Delta t_{AB} = 48 \pm 2$ days, with image A leading. This measurement is crucial for effectively distinguishing microlensing-induced variability from the intrinsic variability of the source quasar. From Figure \ref{lightcurves} we see that the quasar images display considerable time variability, with changes reaching up to $\sim1$ magnitude within periods of approximately two years. Images A and B exhibit parallel variations over long timescales, indicating that these long-term fluctuations primarily originate from the distant quasar itself, suggesting an intrinsic origin (see \citealt{Goicoechea2016}). Furthermore, there is evidence of extrinsic variability (i.e., microlensing) occurring on various timescales, as illustrated by several sharp events in image B that are not observed in image A. 

\section{Data analysis}\label{3}
\subsection{Core ratio measurements}
The cores of emission lines presumably originate from spatially very extended regions (narrow-line region and outer parts of the BLR) that remain largely unaffected by microlensing and intrinsic variability, as demonstrated in studies by \citet{Guerras2013} and \citet{Fian2018blr}. The methodology of quantifying the flux ratios of the line cores between lensed images presents a valid approach for establishing a baseline that is relatively free of microlensing effects. This baseline is essential for accurately assessing (microlensing-based) sizes of emitting regions in quasars. Our study focuses on analyzing three high-ionization lines (including Ly$\alpha$, Si IV, and C IV) alongside two low-ionization lines (C III] and Mg II). For each image and emission line, we remove the continuum by fitting and then subtracting a straight line from the continuum on both sides of the emission line. Given the variable widths of the emission lines, we employ windows of different widths to estimate the continuum for each line and quasar image, deliberately avoiding areas with known emission features. We then define the core fluxes using a narrow interval spanning between $\pm$ 5 to 8 \AA\ centered around the lines' peaks (i.e., $\lambda$1216, $\lambda$1397, $\lambda$1549, $\lambda$1909, and $\lambda$2798). In Table \ref{core_flux_ratios}, we present the average core flux ratios ($\pm1\sigma$ standard deviation) between images A and B calculated from all spectra. Notably, our inferred core flux ratios are in excellent agreement with the values reported by \citet{Shalyapin2014} and \citet{Goicoechea2016}. While it is presumed that the central features of the lines are dominated by photons arising from the narrow-line emission region and the outer parts of the BLR, some of them may also contain a significant number of photons emitted by clouds within the inner zones of the BLR. These clouds are moving almost perpendicularly to the line of sight, exhibiting small projected motions (see, e.g., \citealt{Goicoechea2016}). \citet{Goicoechea2016} reported no evidence of microlensing effects in the cores of the Mg II lines, a finding corroborated by our study. However, the core of C IV is clearly affected by microlensing, and to a certain extent, microlensing also impacts the cores of the other high-ionization lines studied in this work. 
\begin{table}[h]
	\renewcommand{\arraystretch}{1.1}
	\caption{Average line core flux ratios B/A.}
	\begin{tabu} to 0.49\textwidth {X[c]X[c]X[c]X[c]} 
		\hline
		\hline 
		BEL & This work  & Literature$^{(a)}$ & Literature$^{(b)}$ \\ \hline
		Ly$\alpha$ $\lambda 1216$ & $0.32\pm0.01$ & --- & $0.30\pm0.01$ \\
		Si IV $\lambda 1397$ & $0.31\pm0.06$ & --- & $0.36\pm0.02$\\
		C IV $\lambda 1549$ & $0.43\pm0.08$ &  $0.39\pm0.03$ & $0.37\pm0.01$\\
		C III] $\lambda1909$ & $0.27\pm0.02$ & $0.27\pm0.03$ & $0.28\pm0.01$\\
		Mg II $\lambda 2798$ & $0.26\pm0.04$ & $0.24\pm0.03$ & $0.23\pm0.02$ \\ \hline 
	\end{tabu}\\
	
     \small \scalebox{0.955}[1.0]{ \textbf{Notes.} $^{(a)}$\citealt{Shalyapin2014}. $^{(b)}$\citealt{Goicoechea2016}.}
\label{core_flux_ratios}	
\end{table}
\subsection{Continuum variability in spectroscopic data}
The continuum emission of quasars primarily originates from a nuclear region, although recently several studies (see, e.g., \citealt{Chelouche2019,Korista2019,Fian2023diffuse}) offer a slightly different perspective. Microlensing magnification ratios of the continuum at various wavelengths yield crucial insights into the structure and kinematics of the continuum-emitting region. This is because microlensing is sensitive to the size of the source region, with smaller regions leading to higher magnifications. To accurately assess the impact of microlensing on the continuum, it is essential to distinguish this effect from the influences of macro-lensing magnification, which arises from the lens galaxy's smooth potential, and extinction. While each lensed quasar image shares an intrinsic spectrum, they undergo different levels of extinction. This variation occurs because the light from each image traverses distinct paths through the lens galaxy, encountering different concentrations of dust and gas (\citealt{Falco1999,Motta2002,Munoz2004,Munoz2011}). Moreover, the macro-magnification and the differential extinction between the lensed quasar's components are not influenced by the source size and affect both the continuum and emission line flux ratios (\citealt{Motta2002,Mediavilla2005,Mediavilla2009,Mediavilla2011}). We attempt to correct for these effects by estimating the offsets between the continuum adjacent to the emission lines, $(m_B-m_A)_{cont}$, and the magnitude differences in the cores of the emission lines, $(m_B-m_A)_{core}$, between images A and B, \mbox{$\mu_{cont} = (m_B-m_A)_{cont}-(m_B-m_A)_{core}$}. The cores of the emission lines, originating from material distributed across a broad area (narrow-line region and outer regions of the BLR), are typically extensive enough to remain unaffected by microlensing caused by solar mass objects (\citealt{Guerras2013,Fian2018blr}). It is important to note that the BLR is likely stratified, meaning different atoms or ions are located at varying distances from the nuclear continuum-emitting regions. For instance, there is evidence of a correlation between the degree of ionization and proximity to the continuum source, with high-ionization lines located much closer than low-ionization ones (e.g., \citealt{Krolik1991,Peterson2000,Gaskell2007}). In this stratification model of the BLR, the cores of the high-ionization lines might be subject to microlensing. Therefore, caution is advised when interpreting the measured signals in those lines. 
Given the relatively minor wavelength differences between the line cores and the selected wavelength intervals for the continuum (see \mbox{Table \ref{microlensing_continuum}}), this estimator effectively eliminates the influences of macro-magnification and extinction, as demonstrated by \citet{Guerras2013}. 

In Table \ref{microlensing_continuum}, we present the central values employed for fitting the continuum, along with the average magnitude difference (including the $\pm1\sigma$ standard deviation) at each respective wavelength. We chose image A as the reference image as it exhibits less microlensing variability when compared to image B (see \citealt{Shalyapin2014,Goicoechea2016}). The values of $\mu_{cont}$ indicate that B is amplified relative to A by a factor of approximately $3-6$, with greater amplifications at shorter wavelengths. This is in excellent agreement with the findings of \citet{Shalyapin2014}, who reported a factor of $3-5$, and \citet{Goicoechea2016}, who estimated a mean magnification ratio of $\sim5.7$. The microlensing signal observed in the continuum of SDSS J1339+1310 is notably more pronounced than those typically seen in the majority of lensed quasars (\citealt{Fian2016, Fian2018AD,Fian2021AD,Rojas2020,Cornachione2020,Munoz2022}).
\begin{table}[h]
	\renewcommand{\arraystretch}{1.3}
	\caption{Differential microlensing measurements and corresponding size estimates in the continuum.}
	\begin{tabu} to 0.49\textwidth {X[c]X[c]X[c]X[c]} 
		\hline
		\hline 
		$\lambda_{cont}$ (\AA) & $\Delta \lambda_{cont}$ (\AA) & $\mu_{BA}$ (mag) & \hspace*{-0.5mm}$R_{1/2}$ (lt-days) \\  
		(1) & (2) & (3) & (4) \\ \hline
		$1147$ & $12$ & $-2.0\pm0.1$ & $0.6_{-0.3}^{+0.6}$\\
		$1290$ & $6$ & $-2.0\pm0.1$ & $0.6_{-0.3}^{+0.7}$\\
		$1351$ & $9$ & $-1.9\pm0.5$ & $0.6_{-0.3}^{+0.6}$\\
        $1459$ & $13$ & $-1.7\pm0.4$ & $0.7_{-0.3}^{+0.5}$\\  
        $1703$ & $17$ & $-1.4\pm0.3$ & $0.9_{-0.3}^{+0.7}$\\
		$1802$ & $15$ & $-1.8\pm0.1$ & $0.6_{-0.2}^{+0.5}$\\
		$2000$ & $17$ & $-1.7\pm0.2$ & $0.6_{-0.2}^{+0.5}$\\
		$2658$ & $19$ & $-1.4\pm0.2$ & $1.0_{-0.4}^{+0.9}$\\
		$2859$ & $10$ & $-1.3\pm0.2$ & $1.2_{-0.6}^{+0.9}$\\ \hline  	
	\end{tabu}\\
	
		\small \textbf{Notes.} Col. (1): Average central wavelength of the continuum. Col. (2): Average wavelength window used for computing the magnitude differences in the continuum. Col. (3): Average differential microlensing measurements between the images A and B and scatter observed across different epochs. Col. (4): Continuum-emitting region size estimates.\vspace*{-1pt}
\label{microlensing_continuum}	
\end{table}

\subsection{Continuum variability in photometric data}
Intrinsic variability of the source, combined with the light path time delay between the quasar images, could erroneously mimic anomalies in the optical flux ratios. Photometric monitoring of continuum variability in gravitationally lensed quasars offers the benefit of distinguishing between intrinsic and extrinsic variability, once the time delay is known. To analyze extrinsic variations in the light curves of SDSS J1339+1310, we initially model and subtract the variability intrinsic to the quasar itself. We applied the latest time delay estimates of \citet{Shalyapin2021} to shift the light curves of the lagging image B by $-48$ days. Subsequently, we corrected for the magnitude difference between the images by using the core ratio of the low-ionization line Mg II. 
We note that while the C III] emission line core is proximal to the $r$-band, it is likely contaminated by adjacent emission lines. Consequently, we selected the Mg II line core ($B/A = 1.5\pm0.2$) as a reference, which is assumed to offer an uncontaminated baseline for assessing the true magnification ratio of the images in the absence of microlensing. We performed a single spline fitting on the light curve of image A, which is less susceptible to microlensing, to model the quasar's intrinsic variability. Subsequently, by subtracting this spline fitting from the time-shifted light curve of image B, we constructed a difference light curve in which only the microlensing variability remains. During the decade of photometric monitoring, image B appears to have experienced several microlensing events with timescales of $50-100$ days, as observable in the lower panel of Figure \ref{lightcurves}. While we rely on the spline fit to approximate the intrinsic variability of the source, we note that some microlensing variability in image A is likely still present. However, this is irrelevant in our analysis, as we account for the contributions of microlensing of image A in the simulated microlensing difference histograms.
\subsection{BLR variability measurements} 
To assess the (minimal) size of the BEL emitting regions, we normalize the continuum-subtracted spectra across all images and epochs to match the core of the emission line, determined by the flux within a narrow range centered around the line's peak. Assuming that the line cores serve as a reference that is little affected by microlensing and intrinsic variability, comparing the line wing fluxes between image pairs enables us to estimate the lower limit of the emitting region's size. Distinguishing microlensing from intrinsic variability with certainty requires observations separated by the time delay between images. Nevertheless, given the short time delay between images A and B ($48\pm2$ days), we posit that the impact of intrinsic variability on the observed magnitude differences is minimal. This assumption makes it reasonable to attribute most of the observed \mbox{B--A} magnitude differences in the continuum and BEL wings primarily to microlensing. We estimate the microlensing in the line wings, $\mu_{wing} = (m_B-m_A)_{wing}-(m_B-m_A)_{core}$, on either side of the emission line peak, corresponding to a velocity range of \mbox{$3000-8000$ km s$^{-1}$}. The velocity ranges are marked in \mbox{Figure \ref{mean}} and the differential microlensing measurements are listed in Table \ref{microlensing_BEL}. We exclude calculations in line wings that show absorption features, such as the Ly$\alpha$ forest affecting the blue wing of the Ly$\alpha$ line and the pronounced absorption lines in the blue wing of Si IV. 
\begin{table}[h]
	\tabcolsep=0cm
	\renewcommand{\arraystretch}{1.05}
	\caption{Differential microlensing measurements in the BEL wings.}
	\begin{tabu} to 0.49\textwidth {X[c]X[c]X[c]X[c]} 
		\hline
		\hline 
		BEL & Feature & Window (\AA) & $\mu_{BA}$ (mag) \\  
		(1) & (2) & (3) & (4) \\ \hline
		Ly$\alpha$ $\lambda 1216$ & blue wing & --- & --- \\
		 & red wing & $20$ & $-0.4\pm0.1$ \\ \hline 
		Si IV $\lambda 1397$ & blue wing & --- & --- \\
		 & red wing & $23$ & $-0.4\pm0.3$ \\ \hline
		C IV $\lambda 1549$ & blue wing & $26$ & $-0.0\pm0.1$ \\
		 & red wing & $26$ & $-1.0\pm0.3$ \\ \hline
		C III] $\lambda1909$& blue wing & $32$ & $-0.3\pm0.1$ \\
		 & red wing & $32$ & $-0.6\pm0.4$ \\ \hline
		Mg II $\lambda 2798$ & blue wing & $47$ & $-0.1\pm0.3$ \\
		 & red wing & $47$ & $-0.8$ \\ \hline 
	\end{tabu}\\
		
		\small \textbf{Notes.} Cols. (1)--(2): Emission line and line wing. Col. (3): Wavelength window of the line wing. Col. (4): Average differential microlensing between the images A and B and scatter observed across different epochs.\vspace*{-10pt}
\label{microlensing_BEL}	
\end{table}

Significant microlensing-induced variations are evident in the red wings of all emission lines in image B, as well as in the blue wing of C III]. The findings related to C III] warrant careful interpretation, considering that its blue wing is blended with \mbox{Si III]}, Al III, and Ne III/Si II. It is noteworthy that the microlensing magnification of image B relative to image A achieves remarkable values of $2-3$ in the enhanced red wing of C IV, in excellent agreement with the findings of \citet{Goicoechea2016}.

\subsection{UV Fe III gravitational redshift}
\begin{figure}[h]
\includegraphics[width=0.48\textwidth]{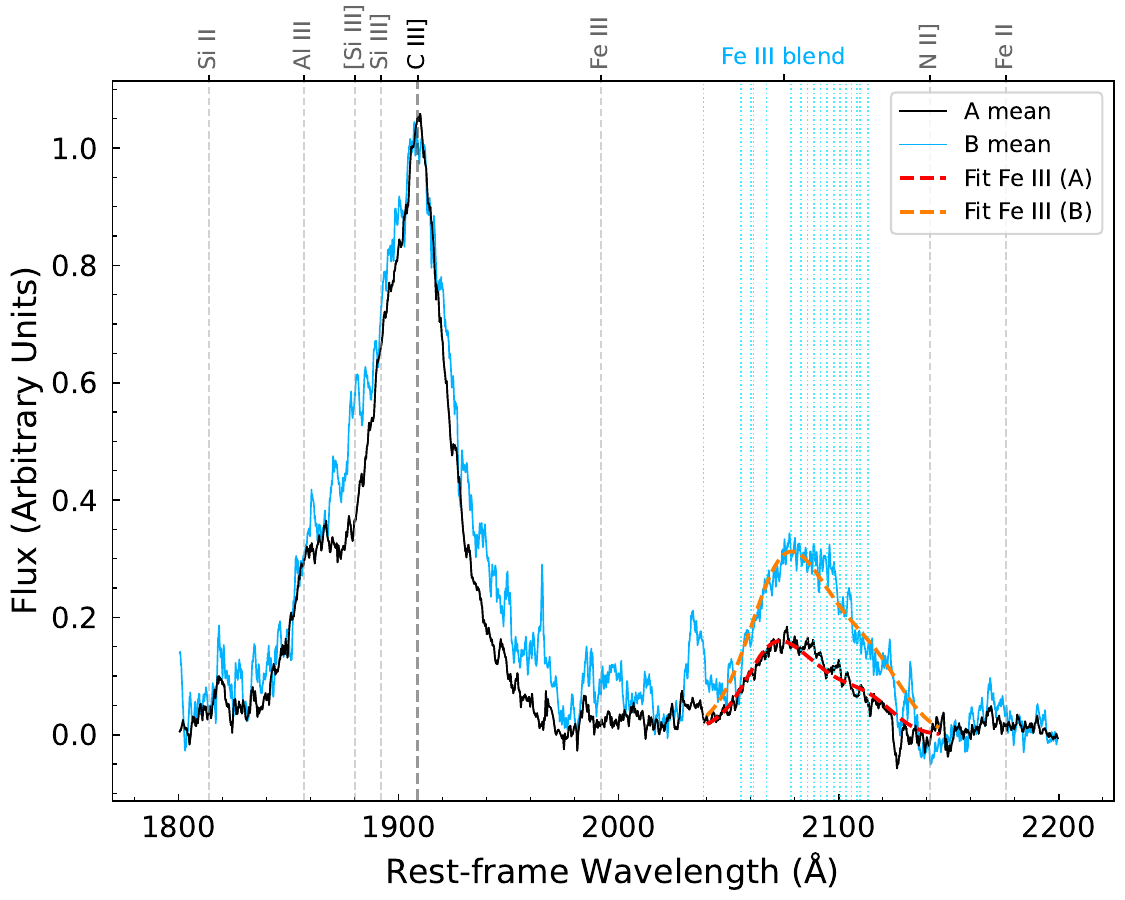}
\caption{Average spectra for images A (in black) and B (in blue), focusing on the regions around the C III] emission line and the \mbox{Fe III $\lambda\lambda$2039-2113} emission feature. The continuum has been subtracted, and the spectra normalized to the core of the C III] line. Vertical dotted blue lines mark the wavelengths of the Fe III lines as per the \citet{Vestergaard2001} template. Dashed red and orange curves represent fits to the \mbox{Fe III $\lambda\lambda$2039-2113} blend in images A and B, respectively, shifted by approximately $\sim8$ \AA\ and $\sim12$ \AA\ to match the template rest-frame. The ordinate is in arbitrary flux units.}
\label{fe3}
\end{figure}

Recently, \citet{Mediavilla2018} observed that the Fe III $\lambda\lambda$2039–2113 feature in quasars is systematically redshifted, a phenomenon that can be explained by the gravitational redshift induced by the central SMBH. This feature is relatively free of contamination by lines from other species, and based on the observed impact of microlensing magnification, it is inferred to originate from an inner region of the BLR (\citealt{Fian2018blr,Fian2021BLR}). We have modeled the \mbox{Fe III $\lambda\lambda2039-2113$} emission feature in the average spectra of images A and B, constructed from five different epochs encompassing the C III] to Mg II wavelength range. First, we fit straight lines to the continuum, one on each side of the C III] emission line and another across two windows flanking the blue and red sides of the Fe III blend. To accommodate variations in the continuum's shape, which can change from epoch to epoch, we have adjusted the windows used for the continuum fitting where necessary. After subtracting the continuum from the spectrum, we normalize the resulting continuum-subtracted spectra to match the core of the C III] line, defined by the flux within a narrow interval centered on the line's peak. The processed spectra for images A and B in the wavelength regions around the C III] line and the \mbox{Fe III} feature are displayed in Figure \ref{fe3}. Significant spectral shape deformations are present in the Fe III $\lambda\lambda$2039–2113 blend redward of C III], where image B appears to be globally affected by high-magnification microlensing. \\

To model the Fe III $\lambda\lambda2039-2113$ blend, we utilized a template comprising 19 distinct Fe III lines with fixed relative amplitudes, as provided by \citet{Vestergaard2001}. Each line in the template is shaped as a Gaussian, broadened uniformly to the same width, defined by $\sigma$ (equivalent to FWHM/2.35). Additionally, the lines are collectively shifted in wavelength by a value of $\Delta\lambda$ and scaled in amplitude by a factor of $K_{scale}$. These three parameters -- $\sigma$, $\Delta\lambda$, and $K_{scale}$ -- are the free variables employed to fit the blend in both images. Figure \ref{fe3} demonstrates that the \citet{Vestergaard2001} template effectively reproduces the observed shape of the Fe III $\lambda\lambda$2039-2113 blend in both images when a redshift of $8.0$ \AA\ for image A and \mbox{$11.7$ \AA}\ for image B, respectively, is applied. The average systematic redshift, \mbox{$\langle \Delta \lambda \rangle = 9.8\pm3.1$ \AA}, aligns well with the findings of \citet{Mediavilla2018}, who reported a value of \mbox{$\langle \Delta \lambda \rangle = 10.3\pm5.9$ \AA}.

To assess the microlensing-induced changes in the UV Fe III emission line blend, we shift the wavelength window $\lambda\lambda2039-2113$ based on the estimated redshift applicable to that specific epoch and image. We then measure the magnitude difference by comparing the fits  to the UV iron blends observed in images A and B within this shifted wavelength range. This analysis yields an average value of $\mu_{Fe III} = 0.8\pm0.2$ mag.
\section{Microlensing simulations}\label{4}
\subsection{Magnification maps}
For simulating the microlensing effects of extended sources, we used the Fast Multipole Method -- Inverse Polygon Mapping (FMM--IPM) algorithm\footnote{\href{https://gloton.ugr.es/microlensing/}{https://gloton.ugr.es/microlensing/}}, as detailed in \citet{Jimenez2022}. This innovative approach integrates the FMM algorithm from \citet{Greengard1987} for computing ray deflections with the IPM algorithm from \citet{Mediavilla2006, Mediavilla2011ipm} for generating magnification maps. Our simulations are based on $3000\times3000$ pixel$^2$ maps, covering an area of approximately $204 \sqrt{M/0.3 M_\odot} \times 204 \sqrt{M/0.3 M_\odot}$ light-days$^2$ on the source plane, where $M$ represents the microlensing mass. For SDSS J1339+1310, the value of the Einstein radius is $R_E = 2.64\times 10^{16} \sqrt{M/0.3 M_\odot}\ \mathrm{cm} = 10.2 \sqrt{M/0.3 M_\odot}$ light-days at the source plane. Our magnification maps for images A and B feature a high resolution of $0.068 \sqrt{M/0.3 M_\odot}$ light-days per pixel, which effectively samples the quasar's optical accretion disk. These maps are defined by two key parameters: the local shear, $\gamma$, and the local convergence, $\kappa$, where $\kappa$ is directly proportional to the surface mass density. The local convergence comprises two distinct components: $\kappa = \kappa_c + \kappa_\star$, where $\kappa_c$ denotes the convergence attributed to continuously distributed matter, such as dark matter, while $\kappa_\star$ corresponds to the convergence caused by stellar-mass point lenses, like microlens stars within the galaxy. The microlenses adhere to a power-law mass function, characterized by a slope similar to that of the Salpeter initial mass function (\citealt{Salpeter1955}). The value of $\alpha\equiv \kappa_\star/\kappa$ represents the fraction of mass in compact objects and quantifies the contribution of stars to the total mass in the lens galaxy. The parameters $\kappa$, $\gamma$, and $\alpha$ for images A and B were taken from \citet{Shalyapin2021} and are listed in Table \ref{macromodel} for reference. We note that \citet{Shalyapin2021} provided ten lens models that match observational constraints with a $\chi^2 \sim 0$. From these, we selected the model that most accurately represents the observed time delay between the images. 
\begin{table}[h]
	\renewcommand{\arraystretch}{1}
	\caption{Macro-model parameters.}
	\begin{tabu} to 0.49\textwidth {X[c]X[c]X[c]X[c]} 
		\hline
		\hline 
		Image & $\kappa$ & $\gamma$ & $\alpha$ \\ 
		(1) & (2) & (3) & (4) \\ \hline
		A & $0.40$ & $0.49$ & $0.28$ \\
		B & $0.63$ & $0.90$ & $0.52$ \\ \hline
	\end{tabu}\\
	
	
	\small \textbf{Notes.} Col. (1): Lensed quasar image. Cols. (2)--(4): Convergence $\kappa$, shear $\gamma$, and fraction of mass in stars $\alpha$ at the quasar image positions. The values were taken from \citet{Shalyapin2021}. 
\label{macromodel}	
\end{table}
\subsection{Source profile}
To model the structure of the unresolved quasar, we adopt circular Gaussian profiles ($I(R)\propto \exp (-{R}^{2}/2r_s^{2})$) to simulate the luminosity distribution of the emitting regions (see, e.g., \citealt{Fian2018blr,Fian2021BLR}). To determine the magnifications for a source of size $r_s$, we convolve the magnification maps with Gaussian profiles having a sigma of $r_s$. The consensus in the field is that the specific shape of the source's emission profile has a minimal impact on microlensing flux variability studies, as demonstrated by \citet{Mortonson2005} and \citet{Munoz2016}. The results are mainly controlled by the half-light radius rather than the detailed intensity profile. Several studies have utilized Gaussian intensity profiles to model the BLR in microlensing studies of quasars (see \citealt{Motta2017,Guerras2013,Sluse2011,Wayth2005}). These profiles are favored due to their simplicity and effectiveness when convolving them with magnification maps. Consequently, we perform convolutions of the maps with Gaussians of varying sizes of $r_s$, spanning an interval between 0.1 and 18 light-days for a mean stellar mass $\langle M \rangle = 0.3 M_\odot$. All linear sizes can be appropriately rescaled for a different mean stellar mass, following the relation $r_s \propto \sqrt{\langle M\rangle}$. The characteristic size $r_s$ is related to the half-light radius by $R_{1/2} = 1.18 r_s$ for Gaussian profiles. After the convolution process, we normalize each magnification map by dividing it by its average value. The resulting histograms of these normalized maps depict the expected microlensing variability. Subsequently, we create the microlensing difference histograms B--A for various $r_s$ values. These simulated microlensing difference histograms are then set for comparison with the observational data, as elaborated in Section \ref{statistics}.

\subsection{Bayesian source size estimation}\label{statistics}
Based on the differential microlensing observed in the wings and adjacent continua of various emission lines between lensed images, we can deduce the size of the emission regions. For this purpose, a statistical approach was employed, treating each microlensing measurement as an individual epoch event. Utilizing all the available observational epochs, we calculate the joint microlensing probability, $P(r_s)$, to derive an average size estimate, following the procedures described in \citet{Guerras2013} and \citet{Fian2018blr,Fian2021BLR}. In the case of the microlensing difference light curve, which provides numerous microlensing measurements, we adopt a more advanced approach. This method entails comparing the microlensing histograms derived from the model, which correspond to convolutions with various source sizes $r_s$, with the histogram of the observed data. For a more detailed explanation of this technique, we refer to previous works such as \citet{Fian2016,Fian2018AD} and \citet{Fian2021AD}. In this study, to accommodate the uncertainties associated with the selected baseline for no microlensing, we permitted the observed data histogram to shift within these uncertainties (see Table \ref{core_flux_ratios}). We have also examined the effects of transverse velocity on the correspondence between the pixels in the simulated magnification map and the observed time span. The light curves utilized cover a span of 10.1 years and comprise 241 data points, with the respective magnification maps displaying a resolution of $0.068 \sqrt{M/0.3 M_\odot}$ light-days per pixel. Assuming an effective transverse velocity of 1000 km s$^{-1}$ and uniform sampling across the light curves, we calculated that approximately 1.3 data points are associated with a single pixel. This does not cause significant alterations to the distribution of the observed histogram.

\section{Results and discussion}\label{5}
Given the sensitivity of microlensing to the size of the source, we use our measurements of microlensing magnification amplitudes to estimate the size of the continuum-emitting region across various wavelengths, as well as the dimensions of the regions emitting the blue and red wings of the BELs and the UV Fe III blend in the lensed quasar SDSS J1339+1310.

\subsection{Continuum-emitting region structure}\label{contcalc}
\begin{figure}[h]
\centering
\includegraphics[width=0.45\textwidth]{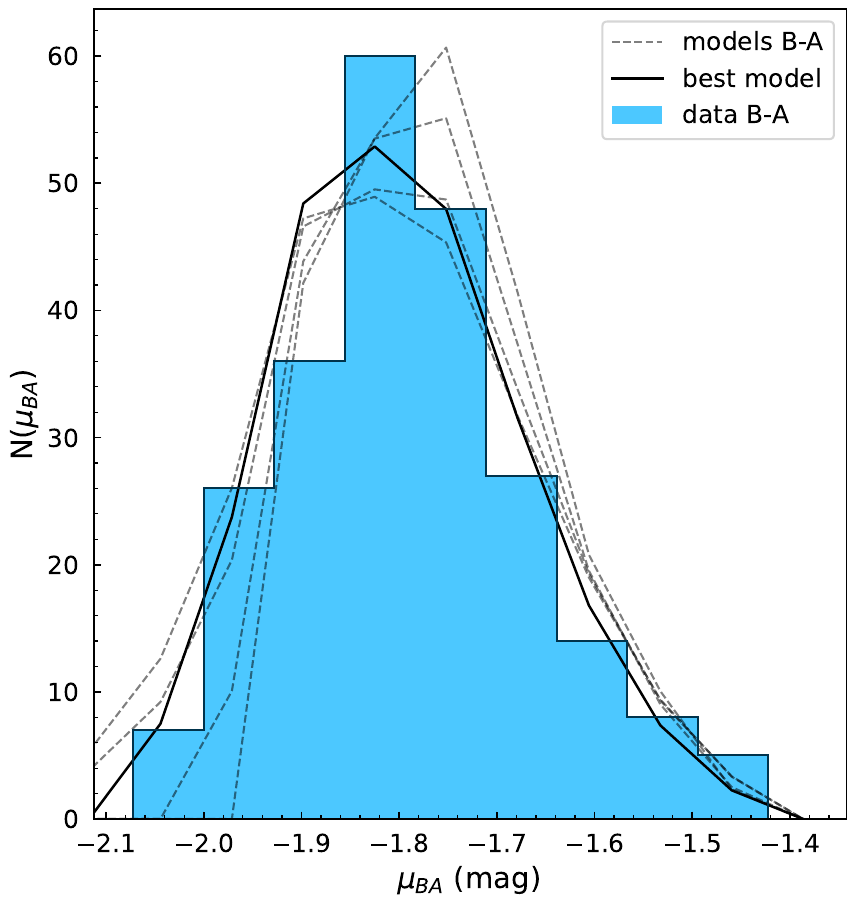}
\caption{Microlensing frequency distributions derived from the observed light curves (illustrated with a blue-colored histogram) compared to those obtained from simulated microlensing magnification maps (depicted by gray and black lines). The polygonal lines represent model histograms for various values of $R_{1/2}$ ranging from 2.0 to 3.0 light-days. The thick line highlights the model's best fit to the observed data, with $R_{1/2} =  2.2\pm0.3$ light-days. We note that the observational histogram was shifted by -0.07 mag to achieve the best alignment with the model histograms.}
\label{histogram}
\end{figure}

We studied the structure of the lensed quasar's continuum-emitting region through the statistics of microlensing magnifications, as inferred from rest-frame UV spectral data and photometric monitoring analyses. Utilizing Bayesian methods, as detailed in Section \ref{statistics}, we estimated the probability of $r_s$ based on the observed differential microlensing between images A and B. Using spectroscopic data, this approach yielded half-light radii for the continuum source ranging from $0.6$ to $1.2$ light-days across the wavelength range of 1150 \AA\ to 2850 \AA\ (see \mbox{Table \ref{microlensing_continuum}}). Analysis of the pronounced microlensing variations in the 2009--2019 \mbox{$r$-band} light curves allowed us to constrain the half-light radius of the UV continuum-emitting region at \mbox{1930 \AA}. By comparing the microlensing difference histogram derived from the data with the simulated microlensing histograms for sources with varying $r_s$ values (see Figure \ref{histogram}), we determined a half-light radius of $R_{1/2} = 2.2\pm0.3$ light-days for the region emitting the $r$-band continuum. We observed that shifting the observational histogram by -0.07 mag yielded the optimal alignment with the model histograms. Our result is in agreement within uncertainties with the recent findings of \citet{Shalyapin2021}, who reported a size of $R_{1/2} = 1.0_{-0.6}^{+7.3}$ light-days. \\

\begin{figure}
\centering
\includegraphics[width=0.45\textwidth]{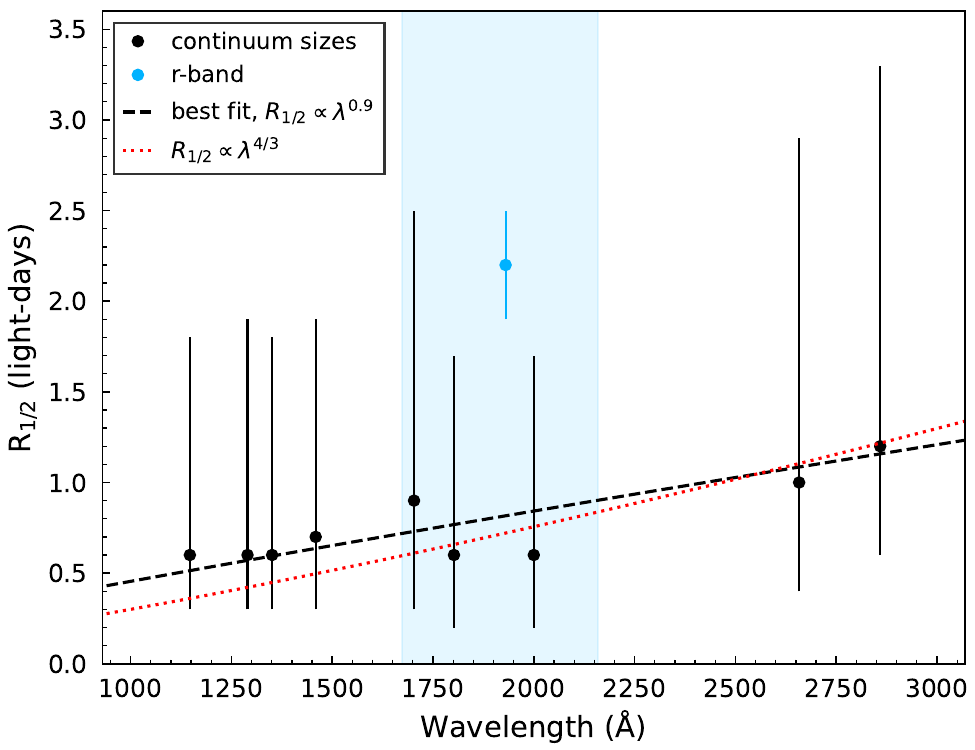}
\caption{Microlensing-based continuum-emitting sizes as a function of wavelength. The dashed black line shows the best fit to the data, with a power-law index of $\beta \sim 0.9$. The red dotted line is a fit with a fixed theoretical power-law index of $\beta = 4/3$, as expected for an optically thick and geometrically thin accretion disk. The blue-colored point corresponds to the $r$-band continuum, and the shaded blue area illustrates the wavelength range covered by the $r$-band.} 
\label{r_lambda}
\end{figure}

Figure \ref{r_lambda} displays the microlensing-based UV continuum-emitting region sizes as a function of wavelength. We fit our size spectrum using a disk model $r_{th}(\lambda) \propto \lambda^{\, \beta}$, where the power-law index $\beta$ -- a marker of the disk's temperature gradient -- is adjustable. From Figure \ref{r_lambda}, it is evident that both the estimated sizes and the physical model roughly align with the expectations of an optically thick geometrically thin accretion disk (see, e.g., \citealt{Shakura1973}), where $\beta = 4/3$. However, our findings suggest a slightly flatter slope than this theoretical model predicts, with the best fit achieved when $\beta= 0.9\pm0.2$. This observation of a flatter size-wavelength relation is consistent with several microlensing campaigns, such as those reported by \citet{Morgan2010}, \citet{Blackburne2011}, \citet{Jimenez2014}, and \citet{Munoz2016}. We note that the $r$-band emitting region, marked by a blue point in Figure \ref{r_lambda}, was excluded from the size spectrum fitting. This decision was based on the fact that this region encompasses a portion of the broad C III] emission line, resulting in a bias towards larger sizes. Additionally, it should be noted that the results regarding the continuum adjacent to the C III] should be interpreted with caution as the presence of (highly variable) \mbox{Fe II} and \mbox{Fe III} lines might influence the continuum blueward of this line.

\subsection{BLR structure} 
\begin{figure}
\includegraphics[width=0.45\textwidth]{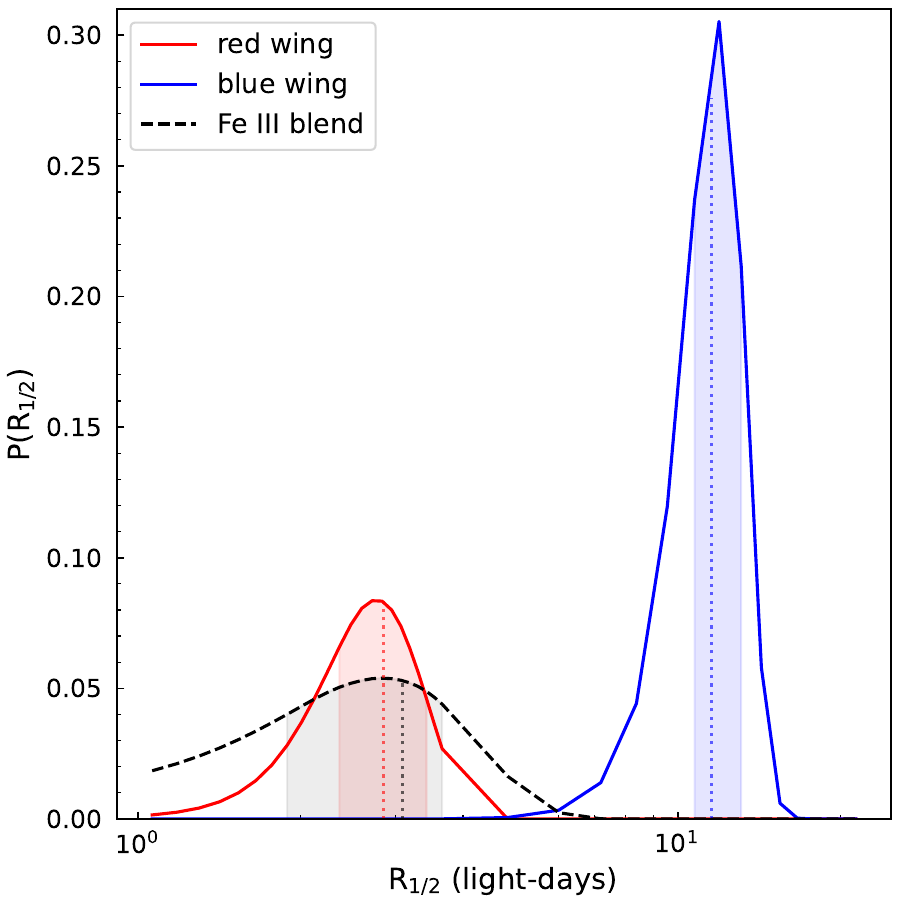}
\caption{PDFs of the half-light radius $R_{1/2}$ emitting the blue and red BEL wings, and the Fe III $\lambda\lambda$2039-2113 blend. The vertical dashed lines indicate the expected size of the emission region, while the shaded regions represent the one-sigma intervals.}
\label{PDFs}
\end{figure}

We applied the methodologies outlined in Section \ref{contcalc} to estimate the dimensions of the continuum-emitting regions, this time focusing on the wings of the BELs and the Fe III emission blend. In contrast to the nuclear continuum emission region, the wings of the BELs originate from a more expanded area located at a greater distance from the central SMBH. We note that due to the limitations of the available spectroscopic observations, we did not estimate the size of the emission regions of the BELs separately. Instead, we provide a joint analysis with rough size estimates for the average emission region sizes of the blue and red line wings. In Figure \ref{PDFs}, we display the probability density functions (PDFs) associated with the regions emitting the blue and red wings of the BELs, as well as the region responsible for the UV Fe III emission. Intriguingly, our analysis reveals a notably small average size for the region emitting the red wing of the BELs ($R_{1/2} = 2.9\pm0.6$ light-days), approximately $30\%$ larger than the size of the $r$-band emitting region. Additionally, the blue BEL wings in this system also originate from a relatively compact area ($R_{1/2} = 11.5\pm1.7$ light-days), smaller than the average BLR size estimates reported by \citet{Guerras2013} and \citet{Fian2018blr,Fian2021BLR} for a sample of lensed quasars (a few tens of light-days). The observed difference between the red and blue wings may be attributed to the velocity field of the red wing residing in a region of higher magnification compared to the blue wing. In single-epoch observations, regions of high magnification yield smaller size estimates, whereas regions of low magnification result in larger size estimates.

Based on the differential microlensing estimates for the UV Fe III blend between images A and B, we deduce that the size of its emission region is $R_{1/2} = 3.1\pm1.2$ light-days across. This indicates that the Fe III feature is formed close to the accretion disk, aligning with the conclusions drawn in studies by \citet{Guerras2013iron}, as well as in our previous works (\citealt{Fian2021BLR,Fian2022_SMBH}).

\subsection{SMBH mass}
Assuming that the redshift of the Fe III $\lambda\lambda$2039-2113 lines is gravitational in origin, we can use the previously determined redshift of the UV Fe III blend together with the inferred microlensing-based size of its emitting region to derive the mass of the central SMBH. This methodology has been successfully applied to diverse quasar samples yielding mass estimates that are statistically compatible with virial masses. Specifically, it was deployed for a sample of ten lensed quasars (\citealt{Mediavilla2018}), extended to another set of ten non-lensed quasars (\citealt{Mediavilla2019}), and employed in a detailed study of \mbox{Q 0957+561} (\citealt{Fian2022_SMBH}). Although this redshift method is generally unaffected by geometry and non-gravitational forces, it is nonetheless reliant on a size estimator expressed as:
\begin{equation}
M_{BH} = \frac{2 c^2}{3 G} \left( \frac{\Delta \lambda}{\lambda} \right)_{\text{Fe III}}{R_{\, \text{Fe III}}},
\label{calculation_mass}
\end{equation}

\noindent where $\Delta \lambda/\lambda$ represents the redshift of the Fe III $\lambda\lambda$2039-2113 emission lines, and $R_{\text{Fe III}}$ denotes the size of the region emitting this blend. Inserting the average redshift, $\langle \Delta\lambda \rangle$, and the half-light radius of the UV Fe III blend, $R_{1/2} = R_{\, \text{Fe III}}$, in Eq. \ref{calculation_mass}, we obtain a SMBH mass of $M_{BH} = (1.7\pm0.5)\times10^8 M_\odot$. This result is in agreement within uncertainties with the virial mass estimate provided by \citet{Shalyapin2021}, $M_{BH} = 4.0_{-2.4}^{+6.0} \times10^8 M_\odot$. The uncertainties in our SMBH mass estimate are proportional to the square root of the combined squared relative errors of the Fe III emitting region size determined via a maximum likelihood method, and the squared standard error of the mean redshift.

\section{Conclusions}\label{6}
SDSS J1339+1339 is, to date, one of the most heavily microlensed quasars, and a substantial amount of photometric monitoring data and spectroscopic observations have been collected over the past decade. In this study, we utilized the newly published macro-model of the lens system by \citet{Shalyapin2021}, alongside cutting-edge techniques to quantitatively model the microlensing statistics (see \citealt{Jimenez2022}). \\

Firstly, it is worth highlighting that SDSS J1339+1339 represents an atypical system, characterized by variations in its light curves occurring over remarkably short timescales. These fluctuations could be attributed to a high transverse velocity and/or a low mass of the microlenses involved. Alternatively, the presence of hotspots or substructures within the accretion disk may also account for these observations. Such features could undergo (de-)magnification if image B crosses one or multiple caustic regions. By analyzing the rapid, microlensing-induced variations in the $r$-band light curves, along with the chromatic changes in the continua adjacent to the BELs, we gained insights into the structure of the continuum-emitting region. The microlensing signals indicate that image B is magnified $\sim 3-6$ times relative to image A at wavelengths between $\sim1000$ and $3000$ \AA, exhibiting a trend of higher amplification at shorter wavelengths. This finding suggests that the sizes of the continuum-emitting regions increase with wavelength, lending support to the hypothesis of disk reprocessing. Additionally, it is noteworthy that the half-light radius we inferred for the region emitting the $r$-band continuum ($R_{1/2} = 2.2\pm0.3$ light-days) is consistent within uncertainties with the value reported by \citet{Shalyapin2021} ($R_{1/2} = 1.0_{-0.6}^{+7.3}$ light-days). However, the size of the emission region in the $r$-band is overestimated by about three times when compared to the sizes obtained from spectroscopic data for the continuum-emitting regions. This overestimation is likely because the $r$-band filter includes part of the BEL C III], which is less affected by microlensing and therefore results in a bias towards a larger emission region size.\\ 

Through the assembly of a dataset comprising seven spectroscopic observations, we have conducted an in-depth analysis of various emission lines, including Ly$\alpha$, Si IV, C IV, C III], and Mg II, as well as the Fe III $\lambda\lambda2039-2113$ emission feature located redward of C III]. In our comparative study of the emission line shapes between images A and B, we have identified noticeable distortions attributable to microlensing. While the blue wings of these emission lines coincide well with each other, a pronounced enhancement is observed in the red wings of image B when compared to that of image A. The observation of microlensing-induced asymmetric deformations in all studied BELs provides strong evidence that the geometry of the BLR is not spherical. Microlensing of a spherically symmetric BLR would lead, in general, to symmetric variations in the emission lines. However, when the BLR is non-spherical, such as a disk-like structure, the microlensing effect is asymmetric, leading to asymmetric variations in the emission lines (\citealt{Schneider1990,Lewis2004,Abajas2002,Sluse2012}). Using Bayesian analysis techniques, we have estimated the average dimensions of the regions responsible for emitting the blue and red wings of the BELs. Our findings reveal that the minimal average size of the region emitting the blue wings is $11.5\pm1.7$ light-days. In contrast, the red wings are emitted from a region approximately four times smaller, measuring $2.9\pm0.6$ light-days. These sizes are significantly smaller compared to the BEL emitting regions typically found in quasars, which, as documented in the literature, are generally within the tens of light-days range (see, e.g., \citealt{Guerras2013,Fian2018blr,Fian2021BLR}). However, similar findings of small sizes for the BLR have also been observed in another system, SDSS J1004+4112, as reported in recent studies by \citet{Hutsemekers2023} and \citet{Fian2023_1004}. \\

We have also deduced that the spectral feature associated with the UV Fe III emission originates from a compact region of $3.1\pm1.2$ light-days across. By combining the microlensing-based size estimate with the measurement of the gravitational redshift for this blend, we were able to calculate the mass of the central SMBH, yielding a value of $M_{BH} = (1.7\pm0.5)\times 10^8 M_\odot$. Our inferred mass is consistent within uncertainties with the virial black hole mass estimate for this system ($M_{BH} = 4.0_{-2.4}^{+6.0}\times 10^8 M_\odot$; see \citealt{Shalyapin2021}), as well as with the average black hole mass determined for a sample of 27 lensed quasars ($M_{BH} = 2.4_{-0.4}^{+1.5} \times 10^8 M_\odot$; see \citealt{Fian2021BLR}). This consistency reinforces the reliability of the (gravitational redshift plus microlensing-based size) mass measurement, making this technique a compelling alternative to other existing methods (e.g., the viral plus reverberation-based size) for measuring black hole masses.\\

\begin{acknowledgements}
We thank the referee James Chan for the valuable comments and suggestions. This research was supported by the grants PID2020-118687GB-C31, PID2020-118687GB-C32, and PID2020-118687GB-C33, financed by the Spanish Ministerio de Ciencia e Innovación. J.A.M. is also supported by the Generalitat Valenciana with the project of excellence Prometeo/2020/085. J.J.V. is also financed by the project FQM-108, financed by Junta de Andalucía. D.C. and S.K. are financially supported by the DFG grant HA3555-14/1 to Tel Aviv University and University of Haifa, and by the Israeli Science Foundation grant no. 2398/19 and 1650/23.
\end{acknowledgements}
\bibliographystyle{aa}
\bibliography{bibliography}

\begin{thebibliography}{61}
\expandafter\ifx\csname natexlab\endcsname\relax\def\natexlab#1{#1}\fi

\bibitem[{{Abajas} {et~al.}(2002){Abajas}, {Mediavilla}, {Mu{\~n}oz},
  {Popovi{\'c}}, \& {Oscoz}}]{Abajas2002}
{Abajas}, C., {Mediavilla}, E., {Mu{\~n}oz}, J.~A., {Popovi{\'c}}, L.~{\v{C}}.,
  \& {Oscoz}, A. 2002, \apj, 576, 640

\bibitem[{{Blackburne} {et~al.}(2011){Blackburne}, {Pooley}, {Rappaport}, \&
  {Schechter}}]{Blackburne2011}
{Blackburne}, J.~A., {Pooley}, D., {Rappaport}, S., \& {Schechter}, P.~L. 2011,
  \apj, 729, 34

\bibitem[{{Chelouche} {et~al.}(2019){Chelouche}, {Pozo Nu{\~n}ez}, \&
  {Kaspi}}]{Chelouche2019}
{Chelouche}, D., {Pozo Nu{\~n}ez}, F., \& {Kaspi}, S. 2019, Nature Astronomy,
  3, 251

\bibitem[{{Cornachione} {et~al.}(2020){Cornachione}, {Morgan}, {Burger},
  {Shalyapin}, {Goicoechea}, {Vrba}, {Dahm}, \& {Tilleman}}]{Cornachione2020}
{Cornachione}, M.~A., {Morgan}, C.~W., {Burger}, H.~R., {et~al.} 2020, \apj,
  905, 7

\bibitem[{{Dawson} {et~al.}(2013){Dawson}, {Schlegel}, {Ahn}, {Anderson},
  {Aubourg}, {Bailey}, {Barkhouser}, {Bautista}, {Beifiori}, {Berlind},
  {Bhardwaj}, {Bizyaev}, {Blake}, {Blanton}, {Blomqvist}, {Bolton}, {Borde},
  {Bovy}, {Brandt}, {Brewington}, {Brinkmann}, {Brown}, {Brownstein}, {Bundy},
  {Busca}, {Carithers}, {Carnero}, {Carr}, {Chen}, {Comparat}, {Connolly},
  {Cope}, {Croft}, {Cuesta}, {da Costa}, {Davenport}, {Delubac}, {de Putter},
  {Dhital}, {Ealet}, {Ebelke}, {Eisenstein}, {Escoffier}, {Fan}, {Filiz Ak},
  {Finley}, {Font-Ribera}, {G{\'e}nova-Santos}, {Gunn}, {Guo}, {Haggard},
  {Hall}, {Hamilton}, {Harris}, {Harris}, {Ho}, {Hogg}, {Holder}, {Honscheid},
  {Huehnerhoff}, {Jordan}, {Jordan}, {Kauffmann}, {Kazin}, {Kirkby}, {Klaene},
  {Kneib}, {Le Goff}, {Lee}, {Long}, {Loomis}, {Lundgren}, {Lupton}, {Maia},
  {Makler}, {Malanushenko}, {Malanushenko}, {Mandelbaum}, {Manera}, {Maraston},
  {Margala}, {Masters}, {McBride}, {McDonald}, {McGreer}, {McMahon}, {Mena},
  {Miralda-Escud{\'e}}, {Montero-Dorta}, {Montesano}, {Muna}, {Myers},
  {Naugle}, {Nichol}, {Noterdaeme}, {Nuza}, {Olmstead}, {Oravetz}, {Oravetz},
  {Owen}, {Padmanabhan}, {Palanque-Delabrouille}, {Pan}, {Parejko},
  {P{\^a}ris}, {Percival}, {P{\'e}rez-Fournon}, {P{\'e}rez-R{\`a}fols},
  {Petitjean}, {Pfaffenberger}, {Pforr}, {Pieri}, {Prada}, {Price-Whelan},
  {Raddick}, {Rebolo}, {Rich}, {Richards}, {Rockosi}, {Roe}, {Ross}, {Ross},
  {Rossi}, {Rubi{\~n}o-Martin}, {Samushia}, {S{\'a}nchez}, {Sayres}, {Schmidt},
  {Schneider}, {Sc{\'o}ccola}, {Seo}, {Shelden}, {Sheldon}, {Shen}, {Shu},
  {Slosar}, {Smee}, {Snedden}, {Stauffer}, {Steele}, {Strauss}, {Streblyanska},
  {Suzuki}, {Swanson}, {Tal}, {Tanaka}, {Thomas}, {Tinker}, {Tojeiro},
  {Tremonti}, {Vargas Maga{\~n}a}, {Verde}, {Viel}, {Wake}, {Watson}, {Weaver},
  {Weinberg}, {Weiner}, {West}, {White}, {Wood-Vasey}, {Yeche}, {Zehavi},
  {Zhao}, \& {Zheng}}]{Dawson2013}
{Dawson}, K.~S., {Schlegel}, D.~J., {Ahn}, C.~P., {et~al.} 2013, \aj, 145, 10

\bibitem[{{Falco} {et~al.}(1999){Falco}, {Impey}, {Kochanek}, {Leh{\'a}r},
  {McLeod}, {Rix}, {Keeton}, {Mu{\~n}oz}, \& {Peng}}]{Falco1999}
{Falco}, E.~E., {Impey}, C.~D., {Kochanek}, C.~S., {et~al.} 1999, \apj, 523,
  617

\bibitem[{{Fian} {et~al.}(2023{\natexlab{a}}){Fian}, {Chelouche}, \&
  {Kaspi}}]{Fian2023diffuse}
{Fian}, C., {Chelouche}, D., \& {Kaspi}, S. 2023{\natexlab{a}}, \aap, 677, A94

\bibitem[{{Fian} {et~al.}(2018{\natexlab{a}}){Fian}, {Guerras}, {Mediavilla},
  {Jim{\'e}nez-Vicente}, {Mu{\~n}oz}, {Falco}, {Motta}, \&
  {Hanslmeier}}]{Fian2018blr}
{Fian}, C., {Guerras}, E., {Mediavilla}, E., {et~al.} 2018{\natexlab{a}}, \apj,
  859, 50

\bibitem[{{Fian} {et~al.}(2016){Fian}, {Mediavilla}, {Hanslmeier}, {Oscoz},
  {Serra-Ricart}, {Mu{\~n}oz}, \& {Jim{\'e}nez-Vicente}}]{Fian2016}
{Fian}, C., {Mediavilla}, E., {Hanslmeier}, A., {et~al.} 2016, \apj, 830, 149

\bibitem[{{Fian} {et~al.}(2021{\natexlab{a}}){Fian}, {Mediavilla},
  {Jim{\'e}nez-Vicente}, {Motta}, {Mu{\~n}oz}, {Chelouche},
  {Gom{\'e}z-Alvarez}, {Rojas}, \& {Hanslmeier}}]{Fian2021AD}
{Fian}, C., {Mediavilla}, E., {Jim{\'e}nez-Vicente}, J., {et~al.}
  2021{\natexlab{a}}, \aap, 654, A70

\bibitem[{{Fian} {et~al.}(2022){Fian}, {Mediavilla}, {Jim{\'e}nez-Vicente},
  {Motta}, {Mu{\~n}oz}, {Chelouche}, \& {Hanslmeier}}]{Fian2022_SMBH}
{Fian}, C., {Mediavilla}, E., {Jim{\'e}nez-Vicente}, J., {et~al.} 2022, \aap,
  667, A67

\bibitem[{{Fian} {et~al.}(2018{\natexlab{b}}){Fian}, {Mediavilla},
  {Jim{\'e}nez-Vicente}, {Mu{\~n}oz}, \& {Hanslmeier}}]{Fian2018AD}
{Fian}, C., {Mediavilla}, E., {Jim{\'e}nez-Vicente}, J., {Mu{\~n}oz}, J.~A., \&
  {Hanslmeier}, A. 2018{\natexlab{b}}, \apj, 869, 132

\bibitem[{{Fian} {et~al.}(2021{\natexlab{b}}){Fian}, {Mediavilla}, {Motta},
  {Jim{\'e}nez-Vicente}, {Mu{\~n}oz}, {Chelouche}, \&
  {Hanslmeier}}]{Fian2021BLR}
{Fian}, C., {Mediavilla}, E., {Motta}, V., {et~al.} 2021{\natexlab{b}}, \aap,
  653, A109

\bibitem[{{Fian} {et~al.}(2024){Fian}, {Mu{\~n}oz}, {For{\'e}s-Toribio},
  {Mediavilla}, {Jim{\'e}nez-Vicente}, {Chelouche}, {Kaspi}, \&
  {Richards}}]{Fian2023_1004}
{Fian}, C., {Mu{\~n}oz}, J.~A., {For{\'e}s-Toribio}, R., {et~al.} 2024, \aap,
  682, A57

\bibitem[{{Fian} {et~al.}(2023{\natexlab{b}}){Fian}, {Mu{\~n}oz}, {Mediavilla},
  {Jim{\'e}nez-Vicente}, {Motta}, {Chelouche}, {Wurzer}, {Hanslmeier}, \&
  {Rojas}}]{Fian2023_0957}
{Fian}, C., {Mu{\~n}oz}, J.~A., {Mediavilla}, E., {et~al.} 2023{\natexlab{b}},
  \aap, 678, A108

\bibitem[{{Fine} {et~al.}(2010){Fine}, {Croom}, {Bland-Hawthorn}, {Pimbblet},
  {Ross}, {Schneider}, \& {Shanks}}]{Fine2010}
{Fine}, S., {Croom}, S.~M., {Bland-Hawthorn}, J., {et~al.} 2010, \mnras, 409,
  591

\bibitem[{{Gaskell} {et~al.}(2007){Gaskell}, {Klimek}, \&
  {Nazarova}}]{Gaskell2007}
{Gaskell}, C.~M., {Klimek}, E.~S., \& {Nazarova}, L.~S. 2007, arXiv e-prints,
  arXiv:0711.1025

\bibitem[{{Gil-Merino} {et~al.}(2018){Gil-Merino}, {Goicoechea}, {Shalyapin},
  \& {Oscoz}}]{Gil-Merino2018}
{Gil-Merino}, R., {Goicoechea}, L.~J., {Shalyapin}, V.~N., \& {Oscoz}, A. 2018,
  \aap, 616, A118

\bibitem[{{Goicoechea} \& {Shalyapin}(2016)}]{Goicoechea2016}
{Goicoechea}, L.~J. \& {Shalyapin}, V.~N. 2016, \aap, 596, A77

\bibitem[{{Greengard} \& {Rokhlin}(1987)}]{Greengard1987}
{Greengard}, L. \& {Rokhlin}, V. 1987, Journal of Computational Physics, 73,
  325

\bibitem[{{Guerras} {et~al.}(2013{\natexlab{a}}){Guerras}, {Mediavilla},
  {Jimenez-Vicente}, {Kochanek}, {Mu{\~n}oz}, {Falco}, \&
  {Motta}}]{Guerras2013}
{Guerras}, E., {Mediavilla}, E., {Jimenez-Vicente}, J., {et~al.}
  2013{\natexlab{a}}, \apj, 764, 160

\bibitem[{{Guerras} {et~al.}(2013{\natexlab{b}}){Guerras}, {Mediavilla},
  {Jimenez-Vicente}, {Kochanek}, {Mu{\~n}oz}, {Falco}, {Motta}, \&
  {Rojas}}]{Guerras2013iron}
{Guerras}, E., {Mediavilla}, E., {Jimenez-Vicente}, J., {et~al.}
  2013{\natexlab{b}}, \apj, 778, 123

\bibitem[{{Hutsem{\'e}kers} \& {Sluse}(2021)}]{Hutsemekers2021}
{Hutsem{\'e}kers}, D. \& {Sluse}, D. 2021, \aap, 654, A155

\bibitem[{{Hutsem{\'e}kers} {et~al.}(2023){Hutsem{\'e}kers}, {Sluse},
  {Savi{\'c}}, \& {Richards}}]{Hutsemekers2023}
{Hutsem{\'e}kers}, D., {Sluse}, D., {Savi{\'c}}, {\DJ}., \& {Richards}, G.~T.
  2023, \aap, 672, A45

\bibitem[{{Inada} {et~al.}(2009){Inada}, {Oguri}, {Shin}, {Kayo}, {Strauss},
  {Morokuma}, {Schneider}, {Becker}, {Bahcall}, \& {York}}]{Inada2009}
{Inada}, N., {Oguri}, M., {Shin}, M.-S., {et~al.} 2009, \aj, 137, 4118

\bibitem[{{Jim{\'e}nez-Vicente} \& {Mediavilla}(2022)}]{Jimenez2022}
{Jim{\'e}nez-Vicente}, J. \& {Mediavilla}, E. 2022, \apj, 941, 80

\bibitem[{{Jim{\'e}nez-Vicente} {et~al.}(2014){Jim{\'e}nez-Vicente},
  {Mediavilla}, {Kochanek}, {Mu{\~n}oz}, {Motta}, {Falco}, \&
  {Mosquera}}]{Jimenez2014}
{Jim{\'e}nez-Vicente}, J., {Mediavilla}, E., {Kochanek}, C.~S., {et~al.} 2014,
  \apj, 783, 47

\bibitem[{{Jim{\'e}nez-Vicente} {et~al.}(2012){Jim{\'e}nez-Vicente},
  {Mediavilla}, {Mu{\~n}oz}, \& {Kochanek}}]{Jimenez2012}
{Jim{\'e}nez-Vicente}, J., {Mediavilla}, E., {Mu{\~n}oz}, J.~A., \& {Kochanek},
  C.~S. 2012, \apj, 751, 106

\bibitem[{{Korista} \& {Goad}(2019)}]{Korista2019}
{Korista}, K.~T. \& {Goad}, M.~R. 2019, \mnras, 489, 5284

\bibitem[{{Krolik} {et~al.}(1991){Krolik}, {Horne}, {Kallman}, {Malkan},
  {Edelson}, \& {Kriss}}]{Krolik1991}
{Krolik}, J.~H., {Horne}, K., {Kallman}, T.~R., {et~al.} 1991, \apj, 371, 541

\bibitem[{{Lewis} \& {Ibata}(2004)}]{Lewis2004}
{Lewis}, G.~F. \& {Ibata}, R.~A. 2004, \mnras, 348, 24

\bibitem[{{Lusso} {et~al.}(2018){Lusso}, {Fumagalli}, {Rafelski}, {Neeleman},
  {Prochaska}, {Hennawi}, {O'Meara}, \& {Theuns}}]{Lusso2018}
{Lusso}, E., {Fumagalli}, M., {Rafelski}, M., {et~al.} 2018, \apj, 860, 41

\bibitem[{{Mediavilla} {et~al.}(2018){Mediavilla}, {Jim{\'e}nez-Vicente},
  {Fian}, {Mu{\~n}oz}, {Falco}, {Motta}, \& {Guerras}}]{Mediavilla2018}
{Mediavilla}, E., {Jim{\'e}nez-Vicente}, J., {Fian}, C., {et~al.} 2018, \apj,
  862, 104

\bibitem[{{Mediavilla} {et~al.}(2019){Mediavilla}, {Jim{\'e}nez-vicente},
  {Mej{\'\i}a-restrepo}, {Motta}, {Falco}, {Mu{\~n}oz}, {Fian}, \&
  {Guerras}}]{Mediavilla2019}
{Mediavilla}, E., {Jim{\'e}nez-vicente}, J., {Mej{\'\i}a-restrepo}, J.,
  {et~al.} 2019, \apj, 880, 96

\bibitem[{{Mediavilla} {et~al.}(2011{\natexlab{a}}){Mediavilla}, {Mediavilla},
  {Mu{\~n}oz}, {Ariza}, {Lopez}, {Gonzalez-Morcillo}, \&
  {Jimenez-Vicente}}]{Mediavilla2011ipm}
{Mediavilla}, E., {Mediavilla}, T., {Mu{\~n}oz}, J.~A., {et~al.}
  2011{\natexlab{a}}, \apj, 741, 42

\bibitem[{{Mediavilla} {et~al.}(2009){Mediavilla}, {Mu{\~n}oz}, {Falco},
  {Motta}, {Guerras}, {Canovas}, {Jean}, {Oscoz}, \&
  {Mosquera}}]{Mediavilla2009}
{Mediavilla}, E., {Mu{\~n}oz}, J.~A., {Falco}, E., {et~al.} 2009, \apj, 706,
  1451

\bibitem[{{Mediavilla} {et~al.}(2005){Mediavilla}, {Mu{\~n}oz}, {Kochanek},
  {Falco}, {Arribas}, \& {Motta}}]{Mediavilla2005}
{Mediavilla}, E., {Mu{\~n}oz}, J.~A., {Kochanek}, C.~S., {et~al.} 2005, \apj,
  619, 749

\bibitem[{{Mediavilla} {et~al.}(2011{\natexlab{b}}){Mediavilla}, {Mu{\~n}oz},
  {Kochanek}, {Guerras}, {Acosta-Pulido}, {Falco}, {Motta}, {Arribas},
  {Manchado}, \& {Mosquera}}]{Mediavilla2011}
{Mediavilla}, E., {Mu{\~n}oz}, J.~A., {Kochanek}, C.~S., {et~al.}
  2011{\natexlab{b}}, \apj, 730, 16

\bibitem[{{Mediavilla} {et~al.}(2006){Mediavilla}, {Mu{\~n}oz}, {Lopez},
  {Mediavilla}, {Abajas}, {Gonzalez-Morcillo}, \&
  {Gil-Merino}}]{Mediavilla2006}
{Mediavilla}, E., {Mu{\~n}oz}, J.~A., {Lopez}, P., {et~al.} 2006, \apj, 653,
  942

\bibitem[{{Morgan} {et~al.}(2010){Morgan}, {Kochanek}, {Morgan}, \&
  {Falco}}]{Morgan2010}
{Morgan}, C.~W., {Kochanek}, C.~S., {Morgan}, N.~D., \& {Falco}, E.~E. 2010,
  \apj, 712, 1129

\bibitem[{{Mortonson} {et~al.}(2005){Mortonson}, {Schechter}, \&
  {Wambsganss}}]{Mortonson2005}
{Mortonson}, M.~J., {Schechter}, P.~L., \& {Wambsganss}, J. 2005, \apj, 628,
  594

\bibitem[{{Motta} {et~al.}(2012){Motta}, {Mediavilla}, {Falco}, \&
  {Mu{\~n}oz}}]{Motta2012}
{Motta}, V., {Mediavilla}, E., {Falco}, E., \& {Mu{\~n}oz}, J.~A. 2012, \apj,
  755, 82

\bibitem[{{Motta} {et~al.}(2002){Motta}, {Mediavilla}, {Mu{\~n}oz}, {Falco},
  {Kochanek}, {Arribas}, {Garc{\'{\i}}a-Lorenzo}, {Oscoz}, \&
  {Serra-Ricart}}]{Motta2002}
{Motta}, V., {Mediavilla}, E., {Mu{\~n}oz}, J.~A., {et~al.} 2002, \apj, 574,
  719

\bibitem[{{Motta} {et~al.}(2017){Motta}, {Mediavilla}, {Rojas}, {Falco},
  {Jim{\'e}nez-Vicente}, \& {Mu{\~n}oz}}]{Motta2017}
{Motta}, V., {Mediavilla}, E., {Rojas}, K., {et~al.} 2017, \apj, 835, 132

\bibitem[{{Mu{\~n}oz} {et~al.}(2004){Mu{\~n}oz}, {Falco}, {Kochanek}, {McLeod},
  \& {Mediavilla}}]{Munoz2004}
{Mu{\~n}oz}, J.~A., {Falco}, E.~E., {Kochanek}, C.~S., {McLeod}, B.~A., \&
  {Mediavilla}, E. 2004, \apj, 605, 614

\bibitem[{{Mu{\~n}oz} {et~al.}(2022){Mu{\~n}oz}, {Kochanek}, {Fohlmeister},
  {Wambsganss}, {Falco}, \& {For{\'e}s-Toribio}}]{Munoz2022}
{Mu{\~n}oz}, J.~A., {Kochanek}, C.~S., {Fohlmeister}, J., {et~al.} 2022, \apj,
  937, 34

\bibitem[{{Mu{\~n}oz} {et~al.}(2011){Mu{\~n}oz}, {Mediavilla}, {Kochanek},
  {Falco}, \& {Mosquera}}]{Munoz2011}
{Mu{\~n}oz}, J.~A., {Mediavilla}, E., {Kochanek}, C.~S., {Falco}, E.~E., \&
  {Mosquera}, A.~M. 2011, \apj, 742, 67

\bibitem[{{Mu{\~n}oz} {et~al.}(2016){Mu{\~n}oz}, {Vives-Arias}, {Mosquera},
  {Jim{\'e}nez-Vicente}, {Kochanek}, \& {Mediavilla}}]{Munoz2016}
{Mu{\~n}oz}, J.~A., {Vives-Arias}, H., {Mosquera}, A.~M., {et~al.} 2016, \apj,
  817, 155

\bibitem[{{Peterson} \& {Wandel}(2000)}]{Peterson2000}
{Peterson}, B.~M. \& {Wandel}, A. 2000, \apjl, 540, L13

\bibitem[{{Popovi{\'c}} {et~al.}(2020){Popovi{\'c}}, {Afanasiev}, {Moiseev},
  {Smirnova}, {Simi{\'c}}, {Savi{\'c}}, {Mediavilla}, \& {Fian}}]{Popovic2020}
{Popovi{\'c}}, L.~{\v{C}}., {Afanasiev}, V.~L., {Moiseev}, A., {et~al.} 2020,
  \aap, 634, A27

\bibitem[{{Rivera} {et~al.}(2023){Rivera}, {Morgan}, {Florence}, {Dahm},
  {Vrba}, {Tilleman}, {Cornachione}, \& {Falco}}]{Rivera2023}
{Rivera}, A.~B., {Morgan}, C.~W., {Florence}, S.~M., {et~al.} 2023, \apj, 952,
  54

\bibitem[{{Rojas} {et~al.}(2020){Rojas}, {Motta}, {Mediavilla},
  {Jim{\'e}nez-Vicente}, {Falco}, \& {Fian}}]{Rojas2020}
{Rojas}, K., {Motta}, V., {Mediavilla}, E., {et~al.} 2020, \apj, 890, 3

\bibitem[{{Salpeter}(1955)}]{Salpeter1955}
{Salpeter}, E.~E. 1955, \apj, 121, 161

\bibitem[{{Schneider} \& {Wambsganss}(1990)}]{Schneider1990}
{Schneider}, P. \& {Wambsganss}, J. 1990, \aap, 237, 42

\bibitem[{{Shakura} \& {Sunyaev}(1973)}]{Shakura1973}
{Shakura}, N.~I. \& {Sunyaev}, R.~A. 1973, \aap, 24, 337

\bibitem[{{Shalyapin} \& {Goicoechea}(2014)}]{Shalyapin2014}
{Shalyapin}, V.~N. \& {Goicoechea}, L.~J. 2014, \aap, 568, A116

\bibitem[{{Shalyapin} {et~al.}(2021){Shalyapin}, {Goicoechea}, {Morgan},
  {Cornachione}, \& {Sergeyev}}]{Shalyapin2021}
{Shalyapin}, V.~N., {Goicoechea}, L.~J., {Morgan}, C.~W., {Cornachione}, M.~A.,
  \& {Sergeyev}, A.~V. 2021, \aap, 646, A165

\bibitem[{{Sluse} {et~al.}(2012){Sluse}, {Hutsem{\'e}kers}, {Courbin},
  {Meylan}, \& {Wambsganss}}]{Sluse2012}
{Sluse}, D., {Hutsem{\'e}kers}, D., {Courbin}, F., {Meylan}, G., \&
  {Wambsganss}, J. 2012, \aap, 544, A62

\bibitem[{{Sluse} {et~al.}(2011){Sluse}, {Schmidt}, {Courbin},
  {Hutsem{\'e}kers}, {Meylan}, {Eigenbrod}, {Anguita}, {Agol}, \&
  {Wambsganss}}]{Sluse2011}
{Sluse}, D., {Schmidt}, R., {Courbin}, F., {et~al.} 2011, \aap, 528, A100

\bibitem[{{Vestergaard} \& {Wilkes}(2001)}]{Vestergaard2001}
{Vestergaard}, M. \& {Wilkes}, B.~J. 2001, \apjs, 134, 1

\bibitem[{{Wayth} {et~al.}(2005){Wayth}, {O'Dowd}, \& {Webster}}]{Wayth2005}
{Wayth}, R.~B., {O'Dowd}, M., \& {Webster}, R.~L. 2005, \mnras, 359, 561

\end{thebibliography}

\end{document}